\pdfoutput=1
\documentclass[ahp]{essaibirk}
\usepackage{graphicx}
\usepackage[applemac]{inputenc} 
\usepackage[french]{babel}
\usepackage{hyperref}
\usepackage{laskar}

\def\ps2{\Frac{\pi}{2}}

\def\llabel{\label} 

\def\etal{{\it et al.}}

\def\Frac#1#2{{{\displaystyle\strut#1}\over{\displaystyle\strut#2}}}

\def\EQM#1{\vcenter{\normalbaselines\m@th
    \ialign{${\displaystyle ##}$\hfil&&\ ${\displaystyle ##}$\hfil\crcr
    \mathstrut\crcr\noalign{\kern-\baselineskip}
    \noalign{\smallskip}
    #1\crcr\mathstrut\crcr\noalign{\kern-\baselineskip}}}}
\def\crm{\cr\noalign{\medskip}}
\def\m@th{\mathsurround=0pt}
\def\text#1{\hbox{#1}}

\newcommand{\be}{\begin{equation}}
\newcommand{\ee}{\end{equation}}

\def\bq{\medskip\hskip 1cm\begin{minipage}{12cm} \normalsize} 
\def\eq{
\end{minipage}\\ \medskip\noindent} 
\def\bqs{\medskip\hskip 1cm\begin{minipage}{12cm} \normalsize\it} 

\newcommand\fige{
\begin{figure} 
\begin{center}
 \includegraphics[width=13 cm]{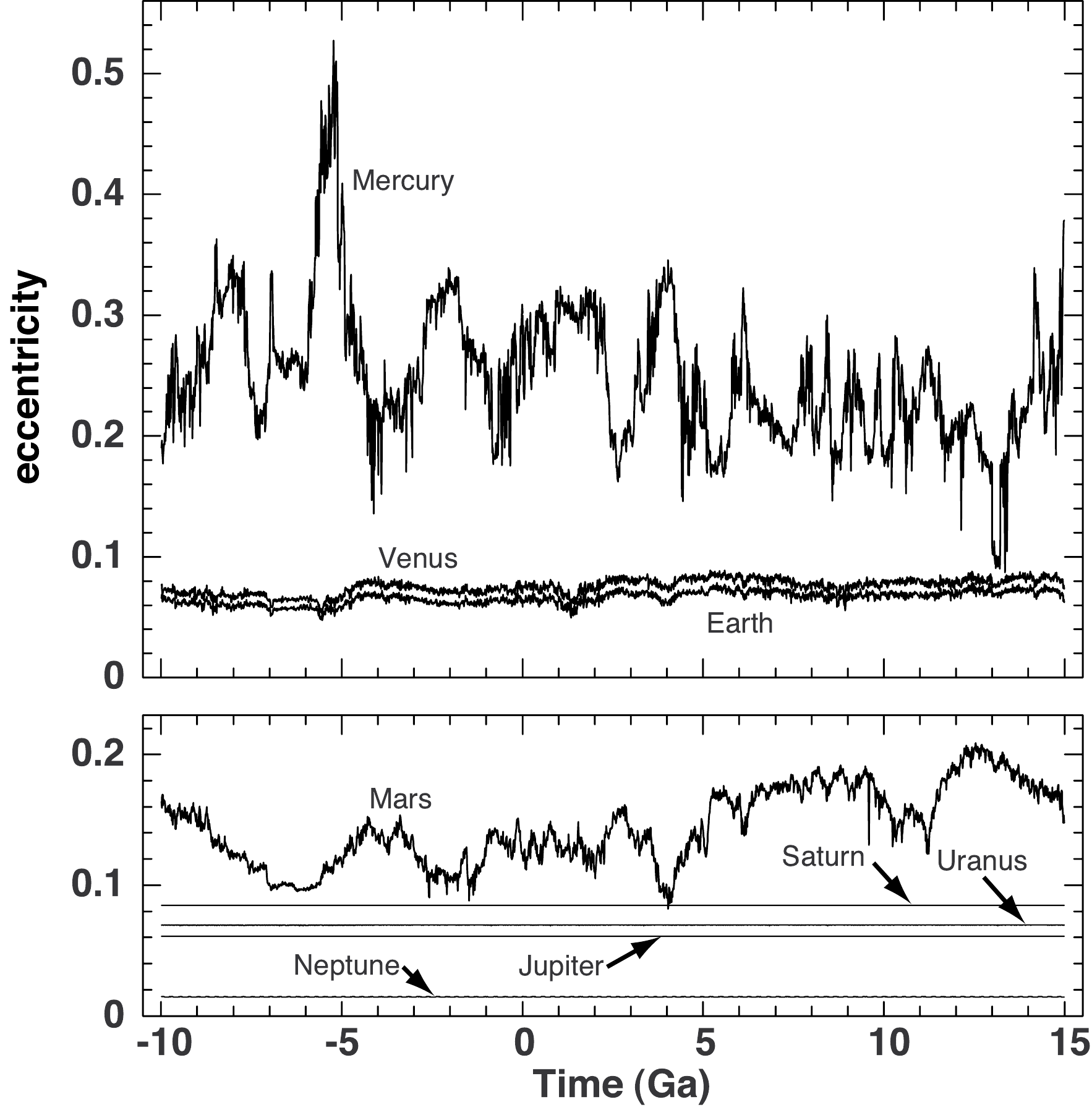}   
\caption{ 
Numerical integration of the averaged equations of the Solar System  from  10 to   15   Gyr.
For each planet, only the maximum eccentricity
reached on slices of 10 Myr is
plotted. The motion of the large planets is very close to a quasiperiodic motion and the amplitude of the 
oscillations of their orbital elements does not vary. Instead, for all inner planets, there is a significant variation of the maximum eccentricity and inclination, which reflects the chaotic diffusion of the  orbits (Laskar, 1994).}
 \label{Fige}
\end{center}
\end{figure}
}

\newcommand\figi{
\begin{figure}[h]  
\begin{center}
 \includegraphics[width=13 cm]{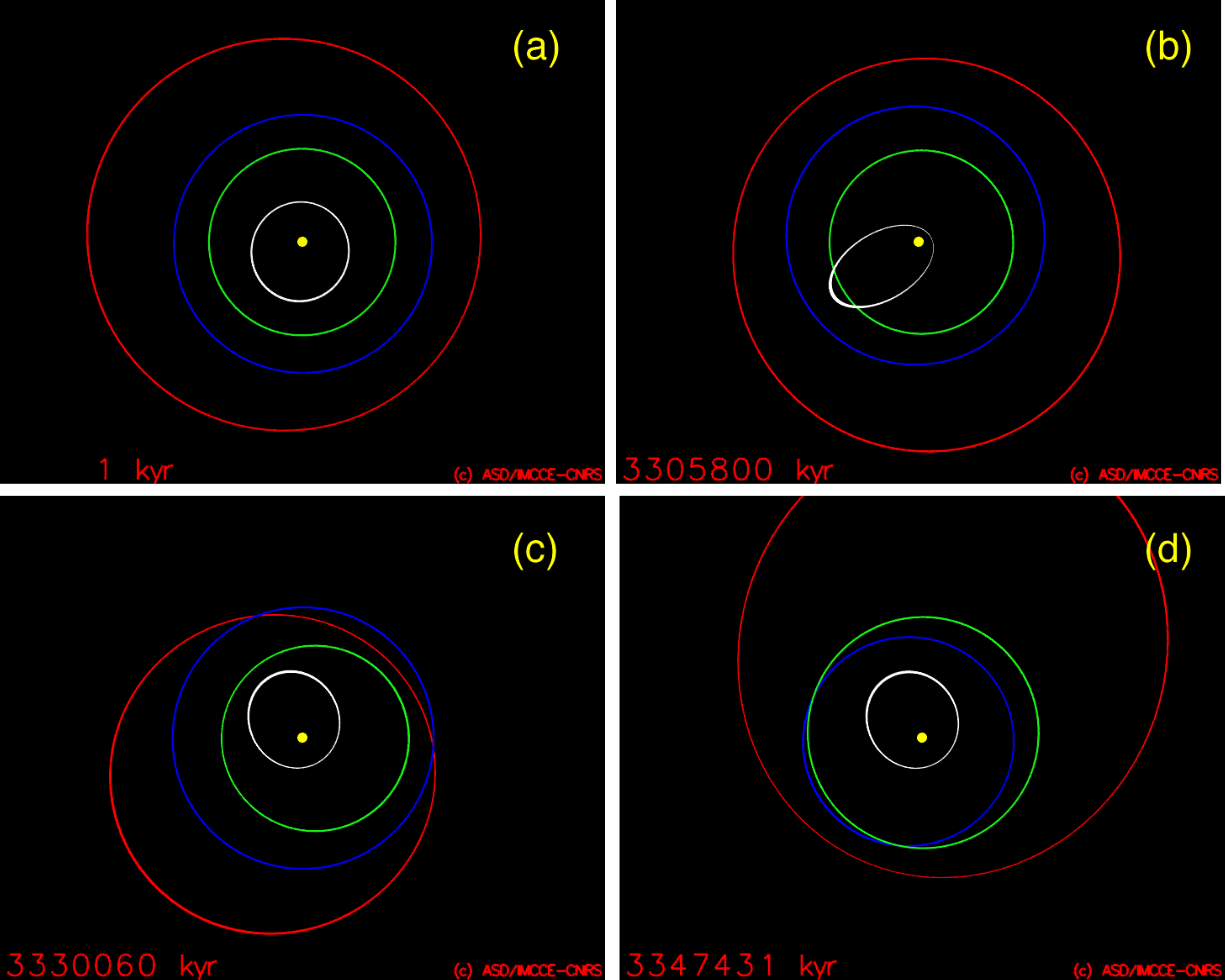}   
\caption{ 
Example of long-term evolution of the orbits of the terrestrial planets:
Mercury (white), Venus (green), Earth (blue), Mars (red).
The time is indicated in thousands of years (kyr).
(a) In the vicinity of the current state, the orbits are deformed under the influence of
planetary perturbations, but without allowing close encounters or collisions.
(b) In about 1 \% of the cases, the orbit of Mercury can deform sufficiently
to allow a collision with Venus or the Sun in less than 5 Gyr.
(c) For one of the trajectories, the eccentricity of Mars increases sufficiently to
allow a close encounter or collision with the Earth.
(d) This leads to a destabilization of the terrestrial planets which also
allow collisions between the  Earth and Venus (Figure adapted from the  results
of the numerical simulations of Laskar and Gastineau, 2009).}
 \label{Figi}
\end{center}
\end{figure}
}

\begin{document}
\Large

\title{Is the Solar System stable?}
\author{Jacques {\sc Laskar}\\
ASD, IMCCE-CNRS UMR8028, Observatoire de Paris, UPMC, \\
       77 avenue Denfert-Rochereau, 75014 Paris, France\\
       laskar@imcce.fr}

\maketitle

\begin{abstract}
Since the formulation of the problem by Newton, and during three centuries, astronomers and mathematicians have sought to demonstrate the stability of the Solar System. Thanks to the numerical experiments of the last two decades, we know now that the motion of the planets in the Solar System is chaotic, which prohibits any accurate prediction of their trajectories beyond a few tens of millions of years. The recent simulations even show that planetary collisions or ejections are possible on a period of less than 5 billion years, before the end of the life of the Sun.
\end{abstract}

\section{
Historical introduction\protect\footnote{  This part is adapted from a lecture of the author on October 19th 2006, at Lombardo Institute (Milan) in honor of Lagrange: 
{\it Lagrange et la stabilité du Système solaire} (Laskar, 2006).}}

Despite the fundamental results of Henri Poincaré about the non-integrability of the three-body problem in the late 19th century, the discovery of the non-regularity of the Solar System’s motion is very recent. It indeed required the possibility of calculating the trajectories  of the planets with a realistic model of the Solar System over very long periods of time, corresponding to the age of the Solar System. This was only made possible in the   past few years. Until then, and this for three centuries, the efforts of astronomers and mathematicians were  devoted to demonstrate the stability of the Solar System.

\subsection{Solar System stability} The problem of the Solar System stability dates back to Newton's statement concerning the law of  gravitation. If we consider a unique planet around the Sun, we retrieve the elliptic motion of Kepler, but as soon as several planets orbit around the Sun, they are subjected to their mutual attraction which disrupts their Kleperian motion. At the end of the volume of Opticks (1717,1730), Newton himself expresses his doubts on this stability which he believes can be compromised by the perturbations of other planets and also of the comets, as it was not known at the time that 
their masses were very small.

\bq And to show that I do not take Gravity for an essential Property of
Bodies, I have added one Question concerning its Cause, chosing to
propose it by way of a Question, because I am not yet satisfied about it
for want of Experiments.

\dots

For while comets move in very excentrick orbs in all manner of
positions, blind fate could never make all the planets move one and the
same way in orbs concentrick, some inconsiderable irregularities
excepted, which may have risen from the mutual actions of comets and
planets upon one another, and which will be apt to increase, till this
system wants a reformation.

\eq

These planetary perturbations are weak because the masses of the  planets in the Solar System are  much smaller than the mass of the Sun (Jupiter's mass is about $1/1000$ of the mass of the Sun). Nevertheless, one may wonder as Newton whether their perturbations could accumulate over very long periods of time and destroy the system. Indeed, one  of the fundamental scientific questions of the 18th century was to first determine if Newton’s law does account in totality for the motion of celestial bodies, and then to know if the stability of the Solar System was granted in spite of the mutual perturbations of planets resulting from this  gravitation law. This problem was even more important  as observations actually showed that Jupiter was getting  closer to the Sun while Saturn was receding  from it. In a  chapter devoted to the {\it secular terms}, De la Lande reports in  the first edition of his "Abrégé d’Astronomie" (1774) the problems that arose from these observations\footnote{Translating  french language of the XVIIIth century is not 
an easy matter, and the translations in english   are provided  here only 
to give a rapid view of the original text. The reader is welcome to propose 
some better translations to the author.}
.

\bq Kepler écrivait en 1625 qu'ayant examiné les observations de
Régiomontanus et de Waltherus,  faites vers 1460 et 1500, il avait
trouvé constamment les lieux de Jupiter \& de Saturne plus ou moins 
avancés qu'ils ne devaient l'être selon les moyens mouvements déterminés
par les anciennes observations de Ptolémée \& celles de Tycho faites
vers 1600. \eq

\bqs
Kepler wrote in 1625, after having  considered the observations of
Regiomontanus and Waltherus made in 1460 and 1500, that he
found consistently that the locations of  Jupiter  \& Saturn were more or less
advanced  as they should be when  their mean motions was determined  
according to  ancient observations of Ptolemy \& those of Tycho made
around 1600.
\eq

Following the work of Le Monnier (1746a, b) which, according to De La
Lande\footnote{De la Lande, Tables Astronomiques de  M. Halley pour les
planètes et les comètes, Paris, 1759}

\bq a démontré le premier, d'une manière suivie et détaillée, après un
travail immense sur les oppositions de Saturne (Mémoire de l'Académie
1746), que non seulement il y a dans cette planète des inégalités
périodiques dépendantes de la situation par rapport à Jupiter, mais que
dans les mêmes configurations qui reviennent après cinquante-neuf ans,
l'erreur des Tables va toujours en croissant. \eq

\bqs
demonstrated for the first time, in a detailed manner,  after
great work on the oppositions of Saturn (Mémoire de l'Académie
1746), that not only  there are some periodic inequalities in this planet 
that  depends on its position relative to Jupiter, but
in the same configurations returning after fifty-nine years,
the error in the 
Tables is  always growing.
\eq

\begin{figure}[!t] \begin{center}
\includegraphics[width=10cm]{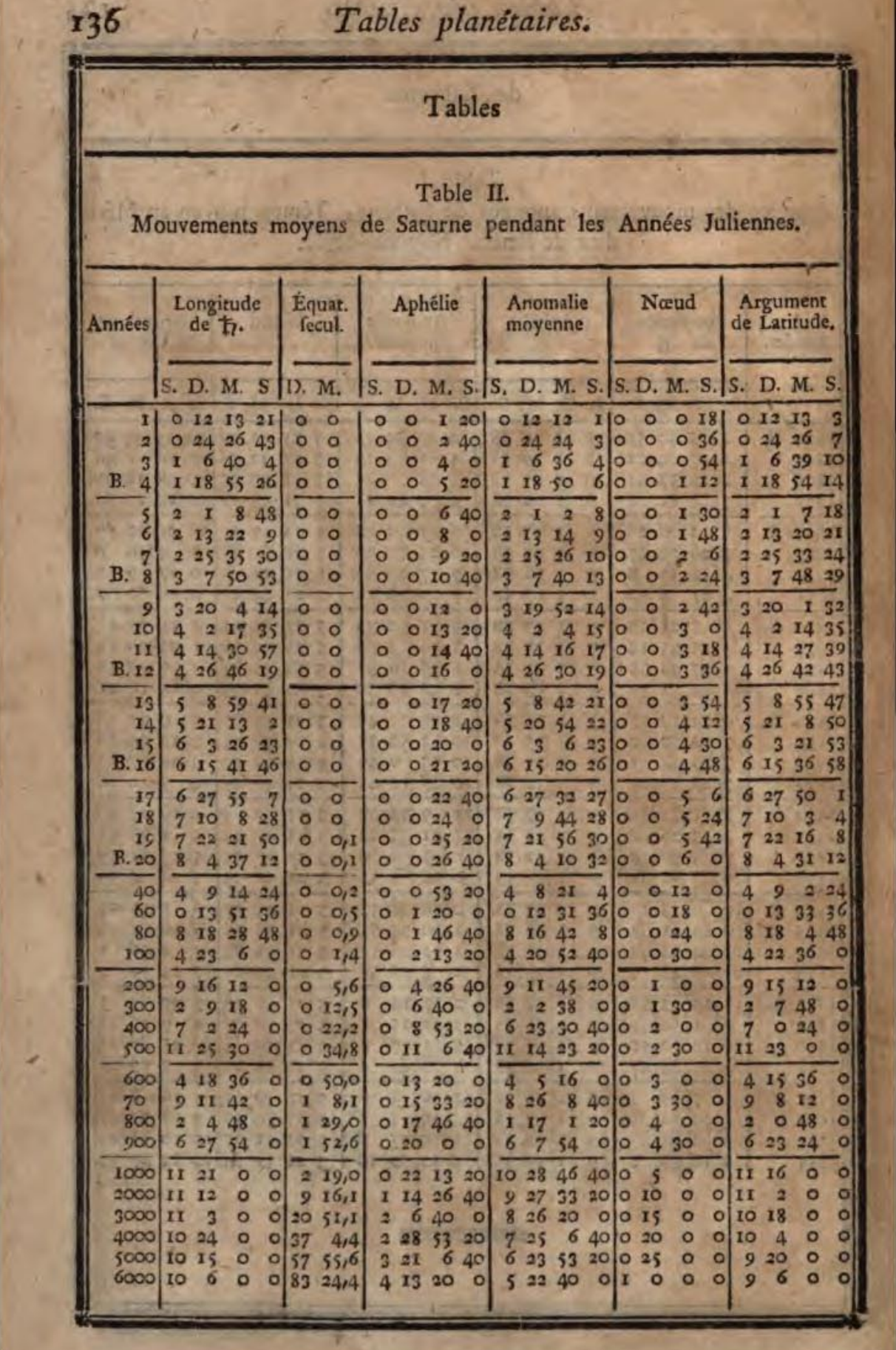} 
\caption{{\small Reproduction of the Tables of Halley {\it Recueil de Tables Astronomiques publié sous la direction de l'Académie Royale des Sciences et Belles-Lettres de Prusse, Vol. II, 1776.}}} 
\llabel{fig_tab}
\end{center} 
\end{figure}

\begin{figure}[!t] \begin{center}
\includegraphics[width=10cm]{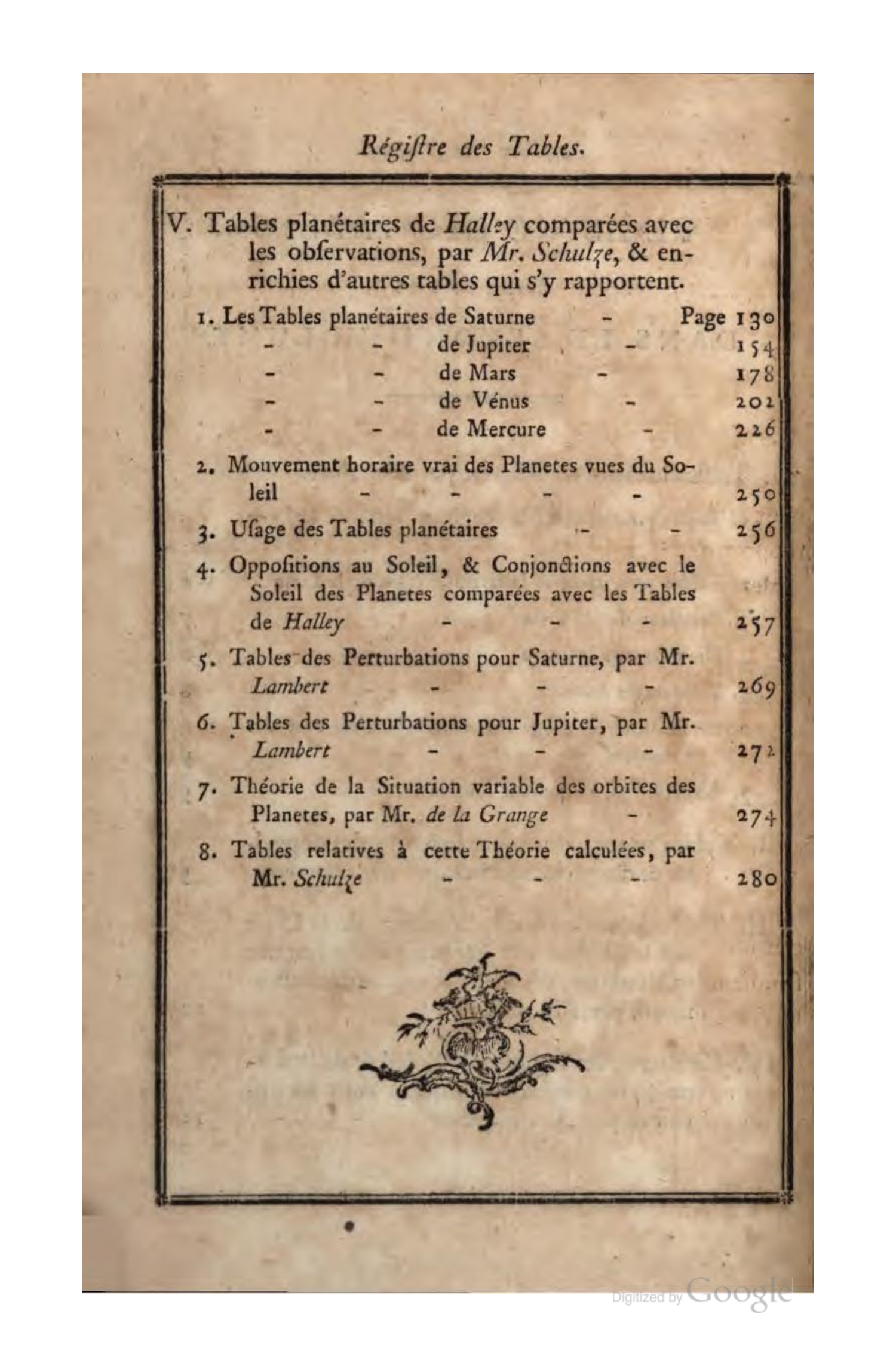} 
\caption{{\small Table of content of {\it Recueil de Tables Astronomiques publié sous la direction de l'Académie Royale des Sciences et Belles-Lettres de Prusse, Vol. II, 1776}. 
It  includes a paper from  Lagrange where he summarizes the results of his memoir on  the nodes and inclinations of the planets
(1774-1778).}}
\llabel{fig_prusse} 
\end{center} \end{figure}

These observations led Halley to introduce a quadratic secular term in the mean longitudes of Jupiter and Saturn. The Tables of Halley became an authority during several decades, and were reproduced in various forms. In particular, by the Royal Academy of Prussia (1776) (Figs. \ref{fig_tab}, \ref{fig_prusse}) during the period when Lagrange lived in Berlin.
These apparent irregularities of Jupiter’s and Saturn’s motions constituted one of the most important scientific  problems of  the 18th century because it was a question of knowing if Newton’s law do account for the motion of planets, and also of deciding on the stability of the Solar System. This led   Paris Academy of Sciences to propose 
several prizes for the resolution of this problem.  Euler was twice awarded 
a Prize of the Academy for these questions, in 1748 and 1752. In his last memoir (Euler, 1752), which laid the foundations of the methods of perturbations, Euler believed that he had demonstrated that Newton’s law induces secular variations in the mean motion of Jupiter and Saturn, variations he found to be of the same sign, contrary to the observations. In reality, we know  now  that  these results from Euler  were  wrong.

\subsection{The 1766 memoir}

\begin{figure}[!t] 
\begin{center}
\includegraphics[width=8cm]{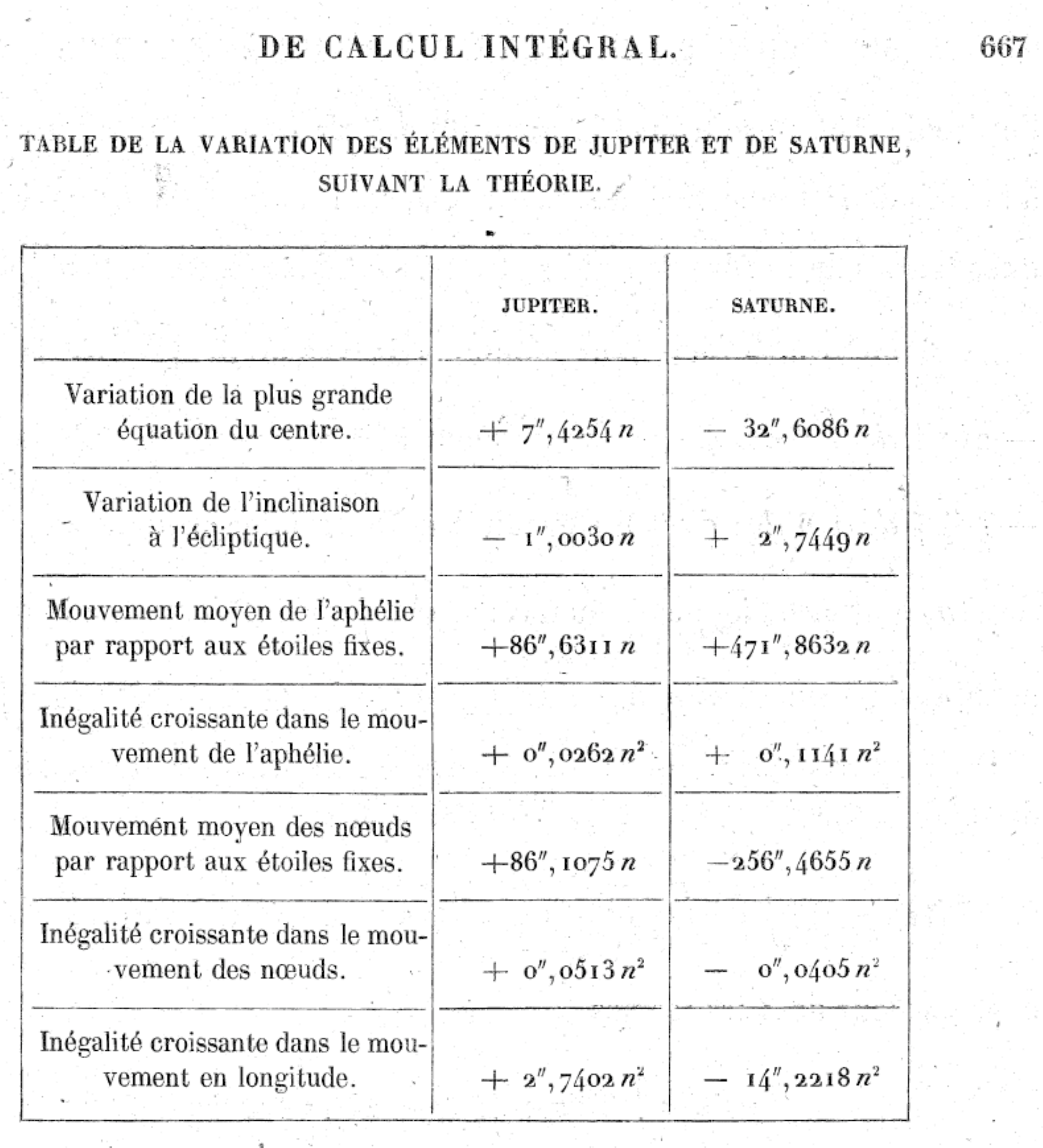} 
\caption{{\small 
Results of Lagrange (1766) for the calculation of the secular inequalities. 
He found the  quadratic term $2".7402 n^2$ in the mean longitude of Jupiter, and  $-14".2218n^2$ in the mean longitude of Saturn, where $n$ is the number of revolutions
of each planet.}}
\llabel{fig_lag1}
\end{center} 
\end{figure}

It is still this important question that Lagrange is trying to solve in his 1766 memoir
{\it Solution de différents problèmes de calcul intégral}, which   appeared in Turin’s memoirs:

\bq 
je me bornerai à examiner ici, d'après les formules données
ci-dessus, les inégalités des mouvements de Jupiter et Saturne qui font
varier l'excentricité et la position de l'aphélie de ces deux planètes,
aussi bien que l'inclinaison et le lieu du n\oe ud de leur orbites, et
qui produisent surtout une altération apparente dans leurs moyens
mouvements, inégalités que les observations ont fait connaître depuis
longtemps, mais que personne jusqu'ici n'a encore entrepris de
déterminer avec toute l'exactitude qu'on peut exiger dans un sujet si
important. 
\eq

\bqs
I will only consider here, according to the above formulas, 
the inequalities in the  motions of Jupiter and Saturn which induce some variations 
in  the eccentricity and the position of aphelion of these two planets,
as well as the inclination and location of the node of their orbits, and mostly
which induce some  apparent alteration in their means
motions,  inequalities that the observations have been made long known, 
but that so far nobody has yet undertaken to 
determine with the  accuracy that is  required for such an important subject.
\eq

It is obvious that Lagrange does not believe in Euler's results. 
However, the care with which Lagrange conducted his own study of the same problem will not be sufficient. Although Lagrange's results are in agreement with the  behavior of the observations (he actually found that Jupiter accelerates while Saturn is slowing down (fig. \ref{fig_lag1}), his calculations are still incorrect. However, this memoir remains a milestone for 
the development of new methods of resolutions of differential equations (see the more detailed work of F. Brechenmacher 2007).

\subsection{The invariance of the semi-major axis}

It is finally Laplace, who will first demonstrate the secular invariance of the semi-major axes of planets, results he publishes in the {\it Memoires de l'Académie des Sciences de Paris}  in 1776. On  the secular inequality of the semi-major axes, he writes:

\bq Elle ne paraît pas cependant avoir été déterminée avec toute la
précision qu'exige son importance. M. Euler, dans sa seconde pièce sur
les irrégularités de Jupiter et de Saturne, la trouve égale pour l'une
et l'autre de ces planètes. Suivant M. de Lagrange, au contraire, dans
le troisième Volume des {\it Mémoires de Turin}, elle est fort
différente pour ces deux corps. \dots j'ai lieu de croire, cependant,
que la formule n'est pas encore exacte. Celle à laquelle je parviens est
fort différente. \dots en substituant ces valeurs dans la formule de
l'équation séculaire, je l'ai trouvée absolument nulle ; d'où je conclus
que l'altération du mouvement moyen de Jupiter, si elle existe, n'est
point due à l'action de Saturne. \eq

\bqs
It does not seem, however, to have been determined with all the
precision required by its importance. Mr. Euler, in his second piece on
the irregularities of Jupiter and Saturn, find it   equal for both  these planets. According to Mr. de Lagrange, on the contrary,
the third volume of {\it Mémoires de Turin}, it is very
different for these two bodies. \dots I have some reasons to believe, however,
that the formula is still not accurate. 
The one which I obtain is 
quite different. \dots by substituting these values in the formula of
the secular equation, I found absolutely zero, from which I conclude
the alteration of the mean motion of Jupiter, if it exists, does 
not  result from the action of Saturn.
\eq

Laplace’s result is admirable, because he succeeds where the most outstanding intellects of the century, Euler and Lagrange, have failed, although they set up  (with d’Alembert) the components which have permitted this discovery. Laplace's result is all the more striking as it runs counter to the observations, which, it is necessary to underline it, had not bothered Euler either. However Laplace does not call into question Newton’s law of gravitation, but makes it necessary to find another cause for the irregularities of Jupiter and of Saturn. Luckily, there is another suitable culprit. Next to the planets, the movement of which seems regular and well ordered, other bodies exist, the comets, of which one had already noticed the very diverse trajectories. As their masses remained unknown at this time, one could evoke their attraction to explain any irregularity in the Solar System.

\bq Il résulte de la théorie précédente que ces variations ne peuvent
être attribuées à l'action mutuelle de ces deux planètes ; mais, si l'on
considère le grand nombre de comètes qui se meuvent autour du Soleil, si
l'on fait ensuite réflexion qu'il est très possible que quelques-unes
d'entre elles aient passé assez près de Jupiter et de Saturne pour
altérer leurs mouvements, \dots il serait donc fort à désirer que le
nombre des comètes, leurs masses et leurs mouvements fussent assez
connus pour que l'on pût déterminer l'effet de leur action sur les
planètes; (Laplace, 1776a). \eq

\bqs
It follows from the above theory that these variations cannot
be attributed to the mutual action of these two planets, but if we
consider  the large number of comets that move around the Sun, if
we then imagine that it is  very possible that some
of them have passed close enough to Jupiter and Saturn to
alter their motions, \dots it would be very desirable that the
number of comets, their masses and their movements were quite
known so that we could determine the effect of their action on the
planets (Laplace, 1776a).\eq

The importance of the analysis of the comets’ trajectories will also be fundamental for the interest that Laplace will take in the study of probability theory. He indeed had to 
discriminate wether the  variety of the  trajectories of comets are the result of chance or not (Laplace, 1776b)

\subsection{Inclinations and eccentricities} 

Laplace had presented his results concerning the invariance of semi-major axes to the Academy in 1773. The following year, in October 1774, Lagrange, then in Berlin, submitted to the Paris Academy of Sciences a new memoir about the secular motions of inclinations and  nodes of the planets. In this memoir, appear for the first time  the linear differential equations with constant coefficients that  represent to the first order the averaged motion of the planetary orbits.

\bq 
Ce Mémoire contient une nouvelle Théorie des mouvements des n\oe uds
et des variations des inclinaisons des orbites des planètes, et
l'application de cette Théorie à l'orbite de chacune des six planètes
principales. On y trouvera des formules générales, par lesquelles on
pourra déterminer dans un temps quelconque la position absolue de ces
orbites, et connaître par conséquent les véritables lois des changements
auxquels les plans de ces orbites sont sujets. 
\eq

\bqs
This Memoir contains a new theory for the motion of the  nodes
and variations of the inclinations of the orbits of the planets, and
the application of this theory to the orbit of each of the six main planets. 
It contains general formulas by which one
can determine, for any time value, the absolute position of any such
orbits, and therefore know the true laws to  which are subject 
the planes of  these orbits.
\eq

One of the important elements in the resolution of these equations is the use of the Cartesian variables

\be 
s=\tan i \sin \Omega \
;\qquad u = \tan i \cos \Omega \ ,
\ee 

where $i$ is the inclination , and  $\Omega$ is the longitude of the node. These variables are almost the same as those which are still used today for the study of planetary motions. Lagrange provided here for the first time a quasi-periodic expression for the motion of the orbital plane 
of the planets  which we can now write in a more synthetic way thanks to complex notation

\be 
u(t)
+ \sqrt{-1} s(t) = \sum_{k=1}^6  \beta_k \exp(\sqrt{-1} s_k t)\ . \ee

The $s_k$ are the eigenvalues of the matrix with constant coefficients of the linear secular System. Of course, Lagrange did not use the matrices  formalism  which will only be put in place much later (see Brechenmacher 2007), but he had to carry out the same computation  of the eigenvalues of a  6 x 6 matrix in an equivalent way. In order to do this, he will proceed by iteration, beginning by the resolution of the Sun-Jupiter-Saturn system. It is impressive to see that in spite of the uncertainties concerning the values of the masses of the inner planets (Mercury, Venus and Mars)\footnote{
Mercury and Venus do not possess  satellites which could provide a good determination of  the masses of the planet 
by applying the third law of Kepler. The satellites of Mars, Phobos and Deimos will only be discovered  many years later, in 1877.}, Lagrange obtained values of the fundamental frequencies of the secular system ($s_k$) that are  very close to the present ones (Tab. \ref{tab.sec}).

\begin{table}[h] \begin{tabular}{ccc} 
\hline 
$k$      & Lagrange (1774)  &   Laskar \etal, 2004  \\ 
\hline 
$s_1$ &     $5.980$&     $5.59$\\
$s_2$ &     $6.311$&     $7.05$\\ 
$s_3$ &     $19.798$&     $18.850$\\
$s_4$ &     $18.308$&     $17.755$\\ 
$s_5$ &     $0$&     $0$\\ 
$s_6$ &    $25.337$&     $26.347$\\ 
\hline 
\end{tabular} 
\caption{\small
Secular frequencies $s_k$  for the motion of the  nodes and
inclinations of the planetary orbits. The values of Lagrange (1774) and
modern values of (Laskar \etal, 2004) are given in arcseconds per
year. It may be surprising that modern values  give less
significant digits than those of Lagrange, but the chaotic  diffusion
in the Solar System causes a significant change in these
frequencies, making it vain for a precise determination of the latter. the
secular zero frequency   $s_5$  results from  the invariance of the angular momentum.
 } 
\llabel{tab.sec} 
\end{table}

At the Academy of Sciences in Paris, Laplace 
was very impressed by the results of Lagrange. He himself had  temporarily left aside his own studies concerning the secular motion of the planetary orbits. He understood immediately the originality and the interest of Lagrange's work and submitted without delay a new memoir to the Academy, concerning the application of Lagrange’s method to the motion of eccentricities and aphelions of planetary orbits (Laplace 1775).

\bq 
Je m'étais proposé depuis longtemps de les intégrer; mais  le peu
d'utilité de ce calcul pour les besoins de l'Astronomie, joint aux
difficultés qu'il présentait, m'avait fait abandonner cette idée, et
j'avoue que je ne l'aurais pas reprise, sans la lecture d'une excellent
mémoire {\it Sur les inégalités séculaires du mouvement des n\oe uds et
de l'inclinaison des orbites des planètes}, que M. de Lagrange vient
d'envoyer à l'Académie, et qui paraîtra  dans un des volumes suivants.
{\it ( Laplace, \oe uvres t VIII, p.355)} 
\eq

\bqs
I had proposed myself, since a long time to incorporate them, but the little
utility of these calculations for the purposes of Astronomy, added to the
difficulties that it presented, made me abandon this idea, and
I confess that I would not have returned to it, without the lecture of  the great
paper {\it ''Sur les inégalités séculaires du mouvement des n\oe uds et
de l'inclinaison des orbites des planètes''} that M. Lagrange has just
send to the Academy, and which will appear in one of the following volumes.
{\it ( Laplace, \oe uvres t VIII, p.355)}
\eq

What is surprising, is that Laplace’s memoir, submitted in December of 1774, is very quickly published, in 1775, with the Academy’s memoirs of 1772, while the original memoir of Lagrange will have to wait until 1778 to be published with the other memoirs of the year 1774. The application to eccentricities and to aphelion is in fact immediate, using the variables

\be l= e \cos \varpi \ ; \qquad h = e \sin \varpi \ . \ee

\bq 
J'ai de plus cherché si l'on ne pourrait pas déterminer d'une
manière analogue les inégalités séculaires de l'excentricité et du
mouvement de l'aphélie, et j'y suis heureusement parvenu; en sorte que
je puis déterminer, non seulement les inégalités séculaires du mouvement
des n\oe uds  et de l'inclinaison des orbites des planètes, les seules
que M. de Lagrange ait considérées, mais encore celles de l'excentricité
et du mouvement des aphélies, et comme j'ai fait valoir que les
inégalités du moyen mouvement et de la distance moyenne sont nulles, on
aura ainsi une théorie complète et rigoureuse de toutes les inégalités
séculaires des orbites des planètes. {\it ( Laplace, \oe uvres t VIII,
p.355)} \eq

\bqs
In addition, I have searched   if one could   determine  
similarly the secular inequalities of eccentricity and
motion of the aphelion, and I happily succeeded, so that
I can determine not only the secular inequalities of the motion
of   nodes and the inclination of the orbits of the planets, the only ones
that were considered by Mr. Lagrange, but also those of the eccentricity
and motion of aphelion, and as I have argued that the
inequalities of the mean motion and the average distance is zero, we
will thus have a complete and rigorous theory of all secular inequalities
of the  orbits of the planets.
{\it ( Laplace, \oe uvres t VIII, p.355)} 
\eq

One may be amazed that  Laplace’s memoir was published before Lagrange’s, and Laplace himself feels obliged to add a note upon this

\bq J'aurai dû naturellement attendre que les recherches de M. de
Lagrange fussent publiées avant que de donner les miennes;  mais, venant
de faire paraître dans les {\it Savants étrangers}, année 1773, un
Mémoire sur cette matière, j'ai cru pouvoir communiquer ici aux
géomètres, en forme de supplément, ce qui lui manquait encore pour être
complet, en rendant d'ailleurs au Mémoire de M. de Lagrange toute la
justice qu'il mérite; je m'y suis d'autant plus volontiers déterminé,
que j'espère qu'ils me sauront gré de leur présenter d'avance l'esquisse
de cet excellent Ouvrage.{\it ( Laplace, \oe uvres t VIII, p.355)} 
\eq

\bqs
I should have naturally waited that the research of Mr.
Lagrange were published before to give mines, but as I just 
published  in {\it  ''Savants étrangers''}, year 1773, a
Memoire on this matter, I thought I could  communicate here
to the geometers,  in the form of a  supplement, which was still lacking 
for
completeness, giving back to the memoire  of Mr. Lagrange all the 
justice it deserves ;
I was even more resolute to do this, as I hope they would be grateful 
to me to present them in advance  the sketch of 
this great work. {\it ( Laplace, \oe uvres t VIII, p.355)} 
\eq

Laplace sends his memoir to Lagrange who sends him back a long letter from Berlin on April, 10 1775:

\bq 
Monsieur et très illustre Confrère, j'ai reçu vos Mémoires, et je
vous suis obligé de m'avoir anticipé le plaisir de les lire. Je me hâte
de vous en remercier, et de vous marquer la satisfaction que leur
lecture m'a donnée. Ce qui m'a le plus intéressé, ce sont vos recherches
sur les inégalités séculaires. Je  m'étais proposé depuis longtemps de
reprendre mon ancien travail sur la théorie de Jupiter et de Saturne, de
le pousser plus loin et de l'appliquer aux autres planètes; j'avais même
dessein d'envoyer à l'Académie un deuxième Mémoire sur les inégalités
séculaires du mouvement de l'aphélie et de l'excentricité des planètes,
dans lequel cette matière serait traitée d'une manière analogue à celle
dont j'ai déterminé les inégalités du mouvement du n\oe ud et des
inclinaisons, et j'en avais déjà préparé les matériaux; mais, comme je
vois que vous avez entrepris vous-même cette recherche, j'y renonce
volontiers, et je vous sais même très  bon gré de me dispenser de ce
travail, persuadé que les sciences ne pourront qu'y gagner beaucoup. 
\eq

\bqs
Mr. and illustrious Colleague, I received your memoirs, and I
am obliged to you to  have anticipated the pleasure of reading it. I look forward
to thank you, and to mark the satisfaction their
reading has given me. What I was most interested in,  are your research
on the secular inequalities. I thought long ago to take back
 my old work on the theory of Jupiter and Saturn, to 
push it further and apply it to other planets.  I even
planned to send a second memory Academy on the  inequalities
of the secular motion of the aphelion and  eccentricity of the planets,
in which the material is treated in a similar manner as what 
I have determined  for the motion of the inequalities of the  node and
inclinations, and I had already prepared the materials, but as I
see that you have undertaken yourself this research, I happily renounce
to it, and I even thanks you for  dispensing me of this
work, convinced that science will  largely  gain   from this.
\eq

So Lagrange specifies that he also had understood that the problem of the eccentricities could be treated in the same way, and because Laplace now deals with this question, Lagrange proposes to give up this subject to him. In fact, this “promise” will not last, and he sends back to d’Alembert a letter dated from May 29, 1775 which shows that he cannot resist continuing his research on this fascinating subject.

\bq 
Je suis près à donner une théorie complète des variations des
éléments des planètes en vertu de leur action mutuelle. Ce que M. de la
Place a fait sur cette matière m'a beaucoup plu, et je me flatte qu'il
ne me saura pas mauvais gré de ne pas tenir l'espèce de promesse que
j'avais faite de la lui abandonner entièrement; je n'ai pas pu résister
à l'envie de m'en occuper de nouveau, mais je ne suis pas moins charmé 
qu'il y travaille aussi de son côté; je suis même fort empressé de lire
ses recherches ultérieures sur ce sujet, mais je le prie de ne m'en rien
communiquer en manuscrit et de ne me les envoyer qu'imprimées; je vous
prie de bien vouloir le lui dire, en lui faisant en même temps mille
compliments de ma part. 
\eq

\bqs
I am ready to give a complete theory for the variations of the 
elements of the planets under their mutual action. That Mr. de la
Place did on this subject I liked, and I flatter myself that
he will not  be offended if I do  not hold the kind of promise that
I made  to completely abandon this subject to him ; I could not resist
to the desire  to look into it again, but I am no less charmed
that he  is also working on it on his side ; I am even very eager to read  
his subsequent research on this topic, but I do ask him not to send me 
any  manuscript and send them to me only in printed form ; I would be obliged that 
you tell him,  with  a thousand compliments from my side.
\eq

Indeed, Lagrange resumed his work and published his results in several memoirs in 1781, 1782, 1783a, b, and 1784 in which he gives the first complete solution of the motion of the six main planets. Perhaps due to the deception he felt following the submission of his article of 1774 to Paris Academy of Sciences, he chose this time to publish his works in the Memoirs of the Academy of Berlin.

\subsection{The great inequality of Jupiter and Saturn{\protect\footnote{A more detailed account of this quest  can be  found in (Wilson, 1985), and (Laskar, 1992)}}}

Laplace had demonstrated the invariance of secular variations of the semi-major axes, considering only the first terms of the expansion of their average perturbations, but the problem of the accordance with  observations remained. He resumed his search with  his theory of Jupiter and Saturn as a base. A first element put him on the right track: the observation of the energy’s conservation in the Sun-Jupiter-Saturn system. If Newton’s law is correct, the conservation of the system’s energy implies that when one of the mean motion  increases, the other must decrease. This is clearly observed. By neglecting the terms of order 2 compared to the masses, he found out that the quantity

\be 
\Frac{m_J}{a_J} + \Frac{m_S}{a_S} 
\ee 
must remain constant.  With Kepler’s law 
($n^2a^3=Cte$)  it gives : 
\be 
\Frac{dn_S}{dn_J} = -\Frac{m_J}{m_S}
\sqrt{\frac{a_J}{a_S}} 
\ee 

where for every planet Jupiter (J) or Saturn (S), $m$ is the mass, $a$ the semi-major axis and $n$ the mean motion. Using the observations  available at the time  (Tab. \ref{tab.mas}), one finds $dn_S/dn_J= -2.32$, which is translated by Laplace as "Saturn deceleration  must be compared to the  acceleration of Jupiter, roughly, as 7 is with 3". Using the values obtained by Halley by comparison with the observations, one obtains $dn_S/dn_J= -2.42$, allowing Laplace to think with great confidence that "the variations observed in the motions of Jupiter and of Saturn result from  their mutual action". Newton’s law does thus not seem to be challenged, but it remained necessary to  find the reason for these variations from Newton's equations. As Laplace demonstrated that there are no secular terms in the first order equations of semi major axes, he inferred that these changes of the average motion of the planets are probably due to short period  terms  (periodic terms with  frequencies that are integer
 combinations of the mean motions  of Jupiter and  Saturn) which would be  of period that is long enough to look like a secular term. A good candidate for this is the term associated to the  combination of longitudes $2 \lambda_J-5\lambda_S$, 
 with period of about 900 years.

\begin{table}[h] \begin{tabular}{cccc} \hline planète     & $1/m$  & 
$a$ (UA) & $n$ ("/365j)  \\ \hline Jupiter &    $1067.195$ &    
$5.20098$  & $ 109182$\\ Saturne&      $3358.40$ &     $9.54007$ & $
43966.5$\\ \hline \end{tabular} \caption{
Values of the parameters of  Jupiter and Saturn used in Laplace's work
 (Laplace, 1785). }
\llabel{tab.mas} \end{table}

This research led Laplace to undertake the construction of a more complete theory of the motion of the Jupiter-Saturn couple. After very long calculations, because to obtain  these terms it is necessary to develop the perturbations to a high degree with respect  to the eccentricities of Jupiter and Saturn, he obtained the following formulas (reduced here to their dominant terms) for the mean longitudes  of Jupiter and Saturn:

\be \EQM{ \lambda_J &= n_J t + \epsilon_J + 20' &\sin (5 n_S t - 2n_S t
+ 49^\circ 8' 40'')\crm \lambda_S &= n_S t + \epsilon_S + 46'50'' &\sin
(5 n_S t - 2n_S t + 49^\circ 8' 40'')\crm } \ee

$\epsilon_J$ and $\epsilon_S$ being the initial conditions for   1700  after J.C. Laplace then corrected the values of the mean motions  of Jupiter and Saturn $n_J$ and $n_S$  with respect  to Halley’s tables. He was then able to compare his new theory, without secular terms in the mean motions (or what is equivalent, without quadratic terms in the mean longitude) to modern and ancient observations. The differences in longitude between his theory and those of new observations (from 1582 to 1786) were all less than $2'$, while the differences with Halley’s tables reached more than $20'$. He also compared his theory with the Chaldean observations of Saturn in 228 BC and of Jupiter at  240 BC transmitted by Ptolemy in the Almageste. These observations are of particularly good quality, because they identify precisely the planet positions  in comparison with known stars. Laplace found a difference with his formulas only $55''$  for the first and $5''$  for the second.

The new theory of the Jupiter-Saturn couple that Laplace completed was therefore  in perfect agreement with the observations from  240 BC to 1715 AD, without the necessity  of an empirical secular term  in the mean motion. The whole theory was entirely derived from  Newton's law of gravitation. Laplace saw here {\it a new proof of the admirable theory of  universal gravity}. He also obtained an important side result, that is, the mass of comets is certainly very small, otherwise their perturbations would have disturbed  Saturn’s orbital motion.

After  this work, the secular terms of the mean motions  will once and for all disappear from the astronomical tables, and in the second edition of his {\it Abrégé d’Astronomie}, De la Lande (1795) will reduce the chapter about the secular equations to a simple paragraph, recalling that Laplace’s calculations on the great inequality of Jupiter and Saturn {\it make the acceleration of the one (Jupiter) and the delaying of the other (Saturn) disappear; their effect is only to make the duration of their revolutions to seem more or less long during nine centuries}.

\subsection{Back to the semi-major axis}

Laplace’s demonstration on the secular invariance of the semi-major axes of the planets 
considered the expansion  up to  degree two in eccentricity and inclinations of the perturbing potential of the planets. Lagrange went back to this problem in 1776 using his method of variations of constants, which allowed him to redo the demonstration, without expansion in eccentricity, and therefore valid for all eccentricities. His demonstration is also particularly simple and very close to the current demonstration. Lagrange will once again come back over this problem in 1808 after Poisson had presented his famous Memoir of about  80 pages (Poisson 1808), where he showed that the invariance of the semi-major axes of the planets is still valid at the second order with respect to  the masses.

Lagrange was in Paris during this time, Member of the Institute, where he had been called by Laplace, in 1787, as a {\it subsidized veteran  of the Academy of Sciences}. In this memoir of 1808, Lagrange showed that using coordinates referring to the barycenter of the Solar System instead of using, as was previously the case, heliocentric coordinates, he succeeded in giving a more symmetric shape to the equations and considerably simplified Poisson’s demonstration.

Indeed, he derived  the differential equations of motion from a single function, and this was the beginning of Lagrangian formalism of variations of constants which already begun to express  its considerable power in this difficult problem. This study conducted to  the general method of Lagrange, described in the {\it Mémoire sur la théorie générale de la variation des constantes arbitraires dans tous les problèmes de la mécanique} (Lagrange 1808 1809). This problem of the Solar System’s stability and of the calculus of the secular terms observed by the astronomers was therefore fundamental in the development of mechanics and perturbative methods and more generally 
in the development of science in the XVIIIth century.

\subsection{The {\it proof}  of stability of Lagrange and Laplace and Le Verrier's question}

After the work of Lagrange and Laplace, the stability of the Solar System seemed to be acquired. The semi-major axes of the orbits had no long-term variations, and their eccentricities and inclinations showed only small variations which did not allow the orbits to intersect and  planets to collide. However, it should be noted that Lagrange and Laplace solutions are very different from  Kepler’s ellipses: the planetary orbits are no longer fixed. They are subject to a double movement of precession with periods ranging from 45000 years to a few million years: precession of the perihelion, which is the slow rotation of the orbit in its plan, and precession of nodes, which is the rotation of the orbital plane in space.

Lagrange and Laplace have written the first {\it proof} of the stability of the Solar System. But as Poincaré (1897) emphasizes in a general audience paper  about the stability of the Solar System:

\bq
Les personnes qui s'intéressent aux progrès de la Mécanique céleste, ... , doivent 
éprouver quelque étonnement en voyant combien de fois on a démontré 
la stabilité du système solaire.

Lagrange l'a établie d'abord, Poisson l'a démontrée de nouveau, 
d'autres démonstrations sont venues depuis, d'autres viendront encore. 
Les démonstrations anciennes étaient-elles insuffisantes, ou sont-ce 
les nouvelles qui sont superflues ?

L'étonnement de ces personnes redoublerait sans doute, si on leur disait 
qu'un jour peut-être un mathématicien fera voir, 
par un raisonnement rigoureux, que le système planétaire est instable.
\eq

\bqs
Those who are interested in the progress of celestial mechanics, ... must
feel some astonishment at seeing how many times  
the stability of the Solar System  has been demonstrated.
 
Lagrange established it  first , Poisson has demonstrated it again,
other demonstrations came afterwards, others will come again.
Were the old demonstrations insufficient, or are  
the new ones  unnecessary?

The astonishment of those people would  probably double, if they would be told
that perhaps one day a mathematician will demonstrate,
by a rigorous reasoning, that the planetary system is unstable.
\eq

In fact, the work of Lagrange and Laplace concerned only the linear approximation of the average motion of the planets. In modern language, we can say they demonstrated that the origin (equivalent to planar circular motions) is an elliptical fixed point in the secular phase space, obtained after averaging of order one with respect to the mean longitudes. Later on, Le Verrier (1840, 1841)  resumed the computations of Lagrange and Laplace. This was before  his discovery of Neptune in 1846, from the analysis of the  irregularities in  the  motion of Uranus. In a first paper (1840) he  computed the secular system for the planets, 
following the previous works of Lagrange and Laplace, with the addition 
of the computation of the change in the solutions resulting from possible new determinations 
of the planetary masses. Soon after (1841), he reviewed the effects of higher order terms in the perturbation series.
He demonstrated that these terms produce significant corrections  to the  linear equations,
and that the calculations of Laplace and Lagrange could not be used for an indefinite period. Le Verrier (1840, 1841) raised the question of the existence of small divisors in the secular system of the inner planets.  This was even more  important as some of the values of the planetary masses  were very imprecise, and  a change of the  mass values could lead to a very small divisor, eventually 
equal to zero. The problem for Le Verrier was then that the terms of third order 
could be larger than the terms of second order, which in his view compromised the 
convergence of the solutions.

\bq
Ces termes acquièrent par l'intégration de très petits diviseurs; 
et ainsi il en résulte, dans les intégrales, des termes dus
à la seconde approximation, et dont les coefficients surpassent
ceux de la première approximation. Si l'on pouvait
répondre de la valeur absolue de ces petits diviseurs, la
conclusion serait simple: la méthode des approximations
successives devrait être rejetée 
\eq

\bqs
Through integration, these terms acquire    very small divisors;
and thus it results in the integrals,  some  terms from 
the second approximation, and which  coefficients exceed
those of the first approximation. If we could
bound the absolute value of these small divisors, the
conclusion would be  simple: the method of successive approximations
 should be rejected
\eq

The indeterminacy of the masses of the inner planets thus did not allow Le Verrier to decide 
of the stability of the system and he could only ask for the mathematicians’ help to solve the problem.

\bq
II paraît donc impossible, par la méthode des approximations
successives, de prononcer si, en vertu des termes de la
seconde approximation, le système composé de Mercure,
Vénus, la Terre et Mars, jouira d'une stabilité indéfinie; et
l'on doit désirer que les géomètres, par l' intégration des
équations différentielles, donnent les moyens de lever cette
difficulté, qui peut très bien ne tenir qu'à la forme.
\eq

\bqs
It seems thus impossible, by the method of successive approximations
 to decide  whether, owing  the terms of the
second approximation,  the system consisting  of Mercury,
Venus, Earth and Mars will enjoy indefinite stability  and
we must desire that geometers, by the integration of the 
differential equations of motion will provide the means to overcome this
difficulty, which may well  result  only from the form.
\eq

\subsection{Poincaré: the geometer's answer}


But Poincaré (1892-99) will give a negative answer to  Le Verrier’s question. To do this, he completely re-thought the methods of celestial mechanics from the work of Hamilton and Jacobi. Poincaré demonstrated that it is not possible to integrate the equations of the movement of  three celestial bodies subject to their mutual gravitational interaction, and that it is impossible to find an analytical solution representing the  planetary motion, valid  over an infinite time interval. In the same way, he concluded that the  perturbations series  used by astronomers to calculate the motion of the planets are not converging on an open  set of initial conditions.

Poincaré therefore showed that the series of the  astronomers are generally divergent. However, he had a high regard for the work of the astronomers of the time, and also pointed out that these divergent series can still be used as a very good approximation for the motion of the planets for some time, which can be long, but not infinite. Poincaré did not seem to think that his results may have great practical importance, if not precisely for the study of the stability of the Solar System.

\bq
Les termes de ces séries, en effet, décroissent d'abord très
rapidement et se mettent ensuite à croître; mais, comme les
astronomes s' arrêtent aux premiers tennes de la série et bien
avant que ces termes aient cessé de décroître, l'approximation
est suffisante pour les besoins de la pratique. La divergence
de ces développements n'aurait d' inconvénients que si
l'on voulait s'en servir pour établir rigoureusement certains
résultats, par exemple la stabilité du système solaire.
\eq

\bqs
The terms of these series, in fact, decrease first very  
quickly and then begin to grow, but as the
Astronomers's stop after the first terms  of the series,   and well
before  these terms have stop to  decrease, the approximation
is sufficient for the practical use. The divergence of 
these expansions  would have some  disadvantages only if
one wanted to use them to rigorously establish some
specific results, as   the stability of the Solar System.
\eq

It should be noted that Poincaré means here stability on infinite time, which is very different from the practical stability of the Solar System, which only makes sense on a time comparable to its life expectancy time interval. Le Verrier had reformulated the question of the stability of the Solar System by pointing out the need to take into account the terms of higher  degree than those considered by Laplace and Lagrange; Poincaré is even more demanding, asking for the convergence of the series:

\bq
Ce résultat aurait été envisagé par Laplace et Lagrange
comme établissant complètement la stabilité du système
solaire. Nous sommes plus difficiles aujourd'hui parce que la
convergence des développements n'est pas démontrée ; ce
résultat n'en est pas moins important.
\eq

\bqs
This result would have been considered by Laplace and Lagrange
as  establishing completely the stability of the Solar System. 
We are more demanding today because the 
convergence of expansions  has not been demonstrated ;  This
result is nevertheless  important.
\eq

Poincaré demonstrated the divergence of the series used by astronomers in their perturbations computations. As usual, he studied a much larger variety of perturbation  series, but apparently not of immediate interest to astronomers, as they required to modify the initial conditions of the planets. However, Poincaré raise some doubts on  the divergence of this type of series:

\bq
Les séries ne pourraient-elles pas, par exemple, converger
quand $x_1^0$ et $x_2^0$ ont été choisis de telle sorte que le rapport
$n_1/n_2$ soit incommensurable, et que son carré soit au contraire
commensurable (ou quand le rapport $n_1/n_2$ est assujetti à une
autre condition analogue à celle que je viens d'énoncer un
peu au hasard) ?
Les raisonnements de ce chapitre ne me permettent pas
d'affirmer que ce fait ne se présentera pas. Tout ce qu'il m' est
permis de dire, c'est qu ' il est fort invraisemblable
\eq

\bqs
The series could they not, for example, converge
when $ x_1 ^ 0 $ and $ x_2 ^ 0 $ have been chosen so that the ratio
$n_1/n_2$ is incommensurable, and its square is instead
commensurable (or when the ratio $n_1/n_2$ is subject to  
another condition similar to that I have enounced 
somewhat randomly)
The arguments of this chapter do not allow me
to say that this  does not exist. All I am 
allowed to say is that this is highly unlikely.
\eq

Half a century later, in line with the work of Poincaré, the Russian mathematician A. N. Kolmogorov actually demonstrated  that these convergent perturbation series  exists.

\subsection{Back to stability}

Kolmogorov (1954) analyzed again the problem of convergence of the perturbation series of celestial mechanics and demonstrated  that for non-degenerated perturbed Hamiltonian systems, close to the non-regular solutions described by Poincaré, there are still regular quasiperiodic trajectories filling  tori in the phase space. This result is not in contradiction with the result of non-integrability of Poincaré, because these tori, parameterized by the action variables, are isolated. This result has been completed by Arnold (1963a) which demonstrated that, for a sufficiently small perturbation, the set of invariant tori foliated by quasi-periodic trajectories is of strictly positive measure, measure that tends to unity when the perturbation tends to zero. Moser (1962) has established the same kind of results for less strong conditions that do not require the analyticity of the Hamiltonian. These theorems are generically called  KAM theorems, and have been used in various fields. Unfortunately, they do not directly apply to the planetary problem that present some proper  degenerescence (the unperturbed Hamiltonian   depends only on the semi-major axis, and not on the other action variables (related to eccentricity and inclination). This led Arnold to extend the proof of the existence of invariant tori, taking into account the phenomenon of degenerescence. He then applied his theorem explicitly to a planar planetary system with two planets, for a semi-major axes’ ratio close to zero, then demonstrating the existence of quasiperiodic trajectories for sufficiently small values of the planetary masses and eccentricities (1963b). This result was later on extended to more general two planets spatial planetary systems (Robutel, 1995). More recently, Féjoz and Herman (2004) have shown the existence of tori of quasiperiodic orbits in a general system of $N$ planets, but this result still requires extremely small planetary masses.

The results of Arnold brought many discussions, indeed, as  the quasiperiodic KAM tori are isolated, an infinitely small variation of the initial conditions will  turn an infinitely  stable quasiperiodic  solution into a chaotic, unstable solution.
Furthermore, as the planetary system has more than two degrees of freedom, none of the KAM tori separates the phase space, leaving the possibility for the chaotic trajectories to travel  great distances in the phase space. This is the diffusion phenomenon highlighted by Arnold.	

In fact, later results showed that in the vicinity of a regular  KAM torus, the diffusion of the trajectories is very slow (Nekhoroshev, 1977, Giorgilli al. 1989, Lochak, 1993, Morbidelli and Giorgilli, 1995), and may be negligible for a very long time, eventually as long as the age of the universe.
Finally, although the masses of the actual planets are much too large for these  results to be applied directly to the Solar System\footnote{The application of Nekhoroshev theorem  for the  stability in finite time of the Solar System was made by Niederman (1996), but required planetary masses of the order of $10^{-13}$ solar mass.}, it is generally assumed that the scope of these mathematical results goes much further than their demonstrated limits, and until recently it was generally accepted that the Solar System is stable, {\it to any reasonable acceptance of this term}.
Over the last twenty years, the problem of the stability of the Solar System has considerably progressed, largely through the assistance provided by computers which allow extensive analytical calculations and numerical integrations of realistic models of the Solar System on durations that are now equivalent to his age.  But  this progress is also due  as well to the understanding of the underlying dynamics, resulting from the development of the theory of dynamical systems since Poincaré.

\section {Numerical computations}

The orbital motion of the planets in the Solar System has a very privileged status. Indeed, it is one of the best modeled problems of physics, and its study may be practically reduced to the study of the behavior of the solutions of the gravitational equations (Newton’s equations supplemented by relativistic corrections) by considering point masses, except in the case of the interactions of the Earth-Moon system. The dissipative effects are also very small, and even if we prefer to take into account the dissipation by tide effect in the Earth-Moon system to obtain a solution as precise as possible for the motion of the Earth (e.g. Laskar et al., 2004), we can very well  ignore the loss of mass of the Sun.
The mathematical complexity of this problem, despite its apparent simplicity (especially if it is limited to the Newtonian interactions between point masses) is daunting, and has been a challenge for mathematicians and astronomers since its formulation three centuries ago. Since the work of  Poincaré, it is also well known that the perturbative methods that were used in the planetary calculations for almost two centuries cannot provide precise approximations of solutions on an infinite time. Furthermore, as indicated above, the rigorous results of stability by Arnold (1963ab) do not apply to realistic planetary systems.

Since the apparition of computers, the numerical integration of the planetary equations has emerged as a simple way to overcome this complexity of the solutions, but this approach has always been limited up to the present by computer technology.
The first long numerical integrations of the Solar System orbits were limited to the outer planets, from Jupiter to Pluto (Cohen et al., 1973, Kinoshita and Nakai, 1984). Indeed, the more the orbital motion of the planets is fast, the more it is difficult to integrate them numerically, because the  required integration step   decreases with the period of the planet. Using  a conventional numerical method, to integrate the orbit of Jupiter, a integration  step size of 40 days is sufficient, whereas a 0.5 days is necessary to integrate the motion  of the entire Solar System including Mercury.
The first integrations of the outer planets system that were performed over  100 Myr and then 210 Myr (Carpino al, 1987, Nobili al, 1989, Applegate et al., 1986) essentially confirmed the stability of the outer planets system, finding quasiperiodic orbits similar to those of Lagrange or Le Verrier. It is only when Sussman and Wisdom (1988) have extended their calculations on 875 Myr that the first signs of instability in the motion of Pluto have appeared, with a Lyapunov time  (the inverse of the Lyapunov exponent) of 20 Myr. But as the mass of Pluto (which is no longer considered as a  {\it planet} since the resolution of the International Astronomical Union in 2006) is very low (1/130 000 000 the solar mass), this instability does not manifest itself by macroscopic instabilities in the remaining part  of the Solar System, which appeared very stable in all these studies.

\subsection{Chaos in the Solar System}

Numerical integrations allow to obtain very accurate solutions for the trajectories of planets, but are limited by the short time step, necessary to achieve  this precision in the case of the complete  Solar System, where it is necessary to take into account the motion of Mercury, and even of the Moon. It should be noted that, until 1991, the only available numerical integration  for a realistic model of the whole Solar System was the numerical integration of the Jet Propulsion Laboratory DE102 (Newhall et al., 1983), calculated over only 44 centuries.

I opted then for a different approach, using analytical perturbation methods, in the spirit of the work of Lagrange, Laplace, and Le Verrier. Indeed, since these pioneering works,  the {\it Bureau des Longitudes}\footnote{ The {\it Bureau des Longitudes}
was founded on  7 messidor an III (june, 25 1795)
to develop astronomy and celestial mechanics. Its founding members
were Laplace, Lagrange, Lalande, Delambre, Méchain, Cassini,
Bougainville, Borda, Buache, Caroché.}, 
always has been the place of development of analytical planetary theories based on the classical perturbation series (Brumberg and Chapront, 1973, Bretagnon, 1974, Duriez, 1979). Implicitly, these studies assumed that the movement of celestial bodies is quasiperiodic and regular. These methods were essentially the same as those which were used by Le Verrier, with the additional help of computers for symbolic calculations. Indeed, these methods can provide very good approximations of the solutions of the planets over thousands of years, but they will not be able to provide answers to the questions of the stability of the Solar System. This difficulty, which has been known since Poincaré is one of the reasons that motivated the direct numerical integration of the equations of motion.

Nevertheless, the results of the KAM theorems suggested the possibility that  classic perturbative solutions could be developed using computer algebra, to find quasiperiodic solutions of the orbital motions in  the Solar System. However, seeking to build such a solution, I realized that the existence of multiple resonances in the averaged system of the inner planets rendered illusory such an approach (Laskar, 1984). This difficulty led me to proceed in two distinct stages:

The first step is the construction of an average system, similar to the systems studied by Lagrange and Laplace. The equations then do not represent the motion of the planets, but the slow deformation of their orbit. This system of equations, obtained by an averaging of order two over the  fast angles (the mean longitudes) thanks to  dedicated computer algebra programs, comprises  153824 polynomial terms. Nevertheless, it can be considered as a simplified system of equations, because its main frequencies are now the frequencies of precession of the orbit of the planets, and not their orbital periods. The complete system can therefore be  numerically integrated with a very big step size of about 500 years. The averaged contributions of the Moon and of  general relativity are added without difficulty and represent just a few additional terms (Laskar, 1985, 1986).

The second step, namely the numerical integration of the average (or secular) system, is then very effective  and could be performed over more than 200 Myr in only a few hours of computation time. The main result of this integration was to reveal that the whole Solar System, and specifically the inner Solar System (Mercury, Venus, Earth and Mars), is chaotic, with a  Lyapunov time of about 5 million years (Laskar, 1989). An error of 15 meters in the initial position of the Earth gives rise to an error of about 150 meters after 10 Ma, but the same error becomes 150 million km after 100 Ma. It is therefore possible to construct precise ephemeris over a period of a few tens of Ma (Laskar et al., 2004, 2011), but it becomes practically impossible to predict the movement of the planets beyond 100 million years.
	
When these results were published, the only possible comparison was a comparison with the planetary ephemeris DE102, over only 44 centuries. This however allowed to be confident about the results, by comparing the derivatives of the averaged solutions at the origin (Laskar, 1986, 1990). At this time, there was no possibility of obtaining similar results by direct numerical integration.

Thanks to the rapid advances in computer industry, just two years later, Quinn et al. (1991) have been able to publish a numerical integration of the whole Solar System, taking into account the effects of  general relativity and of the Moon, over 3 Myr in the past (later complemented by an integration from - 3 Myr to + 3 Myr). The comparison with the secular solution (Laskar, 1990) then showed a very good agreement, and confirmed the existence of secular resonances in the inner Solar System (Laskar \etal, 1992a). Later, using a symplectic integrator that allowed them to use a large step size for the numerical integration of 7.2 days, Sussman and Wisdom (1992) obtained an integration of the Solar System over 100 Myr, which confirmed the value of the Lyapuvov time of approximately 5 Myr for the Solar System.

\begin{figure}
\includegraphics[width=9cm]{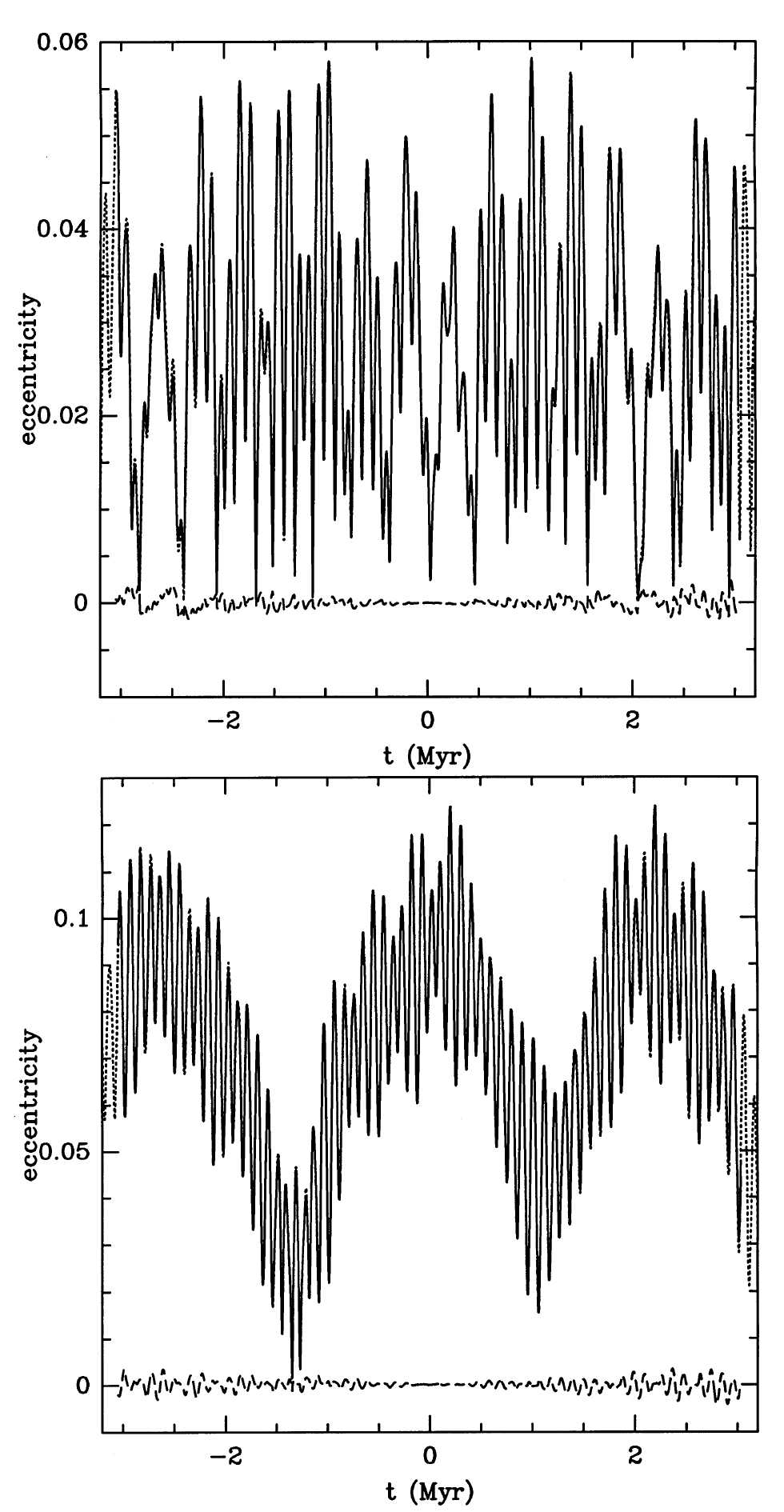}
\caption{ 
The eccentricity of the Earth (a) Mars and (b) from  $-3$Myr to $+3$Myr.
The solid line is the numerical solution from 
(Quinn \etal, 1991), and the dotted line, the secular solution
 (Laskar, 1990). For clarity, the difference between the
two solutions is also plotted (Laskar \etal, 1992).
}   
\llabel{F.QTD90}    
\end{figure}

\subsection{Planetary motions over several million years  (Myr)} 

The variations of eccentricities and inclinations of the planetary orbits are clearly visible on a few Myr (Fig. \ref{F.QTD90}). Over one million years, the solutions resulting from the perturbation  methods of Lagrange and Le Verrier would already give a good estimate of these variations that are essentially due to the linear coupling in the secular equations. On several hundreds of Myr, the behavior of solutions of the external planets (Jupiter, Saturn, Uranus and Neptune) are very similar to the one of the first Myr, and the motion of these planets appears to be very regular, which has also been shown very accurately by   means of frequency analysis (Laskar, 1990).

\subsection{Planetary motions over several billion years  (Gyr)}

Once it is known that the motion of the Solar System is chaotic, with exponential divergence of trajectories that multiplies the error on the initial positions by 10  every 10 Myr, it becomes illusory to try to retrieve, or predict the movement of the planets beyond 100 Myr by the calculation of a single trajectory. However, one can make such a computation to explore the phase space of the system.  The calculated trajectory should then only be considered as a possible trajectory among others after 100 Myr. In (Laskar, 1994), such calculations have even been pushed on several billion years to highlight the impact of the chaotic diffusion of the orbits. In figure \ref{Fige} we no longer represent the eccentricities of the planets, but their maximum value, calculated on  slices of 1 Myr. If the trajectory is quasiperiodic or close to quasiperiodic, this maximum will behave as  a straight horizontal line, corresponding to the sum of the modulus of the amplitudes of the various periodic terms in the quasiperiodic expansion of the solution.

In this way, we are able to eliminate the oscillation of eccentricities resulting from the linear coupling already present in solutions of Lagrange or Le Verrier. The remaining variations of this maximum, which appear in figure  \ref{Figi} are then  the results of only the  chaotic diffusion of the orbits. We see that for all the external planets (Jupiter, Saturn, Uranus, and Neptune), the maximum of the eccentricity is a horizontal line. It means that the motion of these planets is very close to quasiperiodic. However, for all the inner planets, there is a significant chaotic  diffusion  of the eccentricities.
This diffusion is moderate for Venus and the Earth, important for Mars, whose orbit can  reach eccentricities of the order of 0.2 (which does not allow collision with the Earth), and very strong for Mercury which reached an eccentricity of 0.5.
This value is however not sufficient to allow for a collision with Venus, which requires an eccentricity of more than 0.7 for Mercury. But it is well understood that beyond 100 Myr, the trajectories of figure \ref{Figi} only represent a possible trajectory of the Solar System, and a small change in initial conditions will significantly change the trajectories after 100 Myr.

\fige

To find out if collisions between Mercury and Venus are possible, it is therefore necessary to study the variations of the solutions under the influence of a small change in initial conditions. In (Laskar, 1994) I lead this study, using the secular system, showing that it was actually possible to build, section by section, an orbit of collision for Mercury and Venus.
In a first step, the nominal trajectory is integrated over 500 Myr. Then, 4 additional trajectories are integrated, corresponding to small changes of 15 meters of the position reached at 500 Myr. All the trajectories are then integrated over  500 Myr and the trajectory of greater eccentricity of Mercury 
is retained and stopped in the neighborhood of the maximum of eccentricity. This operation  is then repeated and  leads to an eccentricity of Mercury of more than 0.9 in only 13 steps, in less than 3.5 Gyr, allowing thus a collisions with Venus. However, while repeating the same experiments with the trajectories of Venus, Earth and Mars, it was not possible  to construct collisional solutions for these planets.

\subsection{Chaotic diffusion in the Solar System}

The 1994 approach however had some limitations, because the approximation obtained by the averaged equations decreases in accuracy as one approaches the collision. A study using the complete, non- averaged equations was therefore necessary to confirm these results. Despite the considerable increase in the power of  computers since 1994, no complete study of this problem was conducted before 2009.
Actually, because of the chaotic nature of the solutions, the only possible approach is a statistical study of a large number of solutions, with very similar initial conditions. This shows the difficulty of the problem.  Indeed,  before  2009, no direct integration of a single trajectory of the Solar System had yet been published using a realistic model, including the effect of the Moon and general relativity.
To approach this problem, I have firstly carried out such a statistical study, using the averaged equations (whose numerical integration is about 1000 times faster than for the full equations), for 1000 different solutions that were  integrated over 5 Gyr. This  study (Laskar, 2008) showed that the probability to reach very high eccentricities for Mercury ($> 0.6$) is on the order of 1\%.
In this same study, I could also show that over periods of time longer than   500 Myr, the distributions of the eccentricities and inclinations of the inner  planets (Mercury, Venus, Earth, and Mars) followed Rice’s probability densities, and behave like random walks with a very simple empirical distribution law.
These results differed significantly from the results published in 2002 by Ito and Tanikawa, who had integrated 5 orbits on 5 Gyr for a purely (Non-relativistic) Newtonian model.  I therefore also wanted to test the same statistics for a non-relativistic system, thinking that this system would be more stable, such as Ito’s and Tanikawa’s one (2002) who found for Mercury a maximum eccentricity of only 0.35. Much to my surprise, the result, on 1000 numerical solutions  of the secular system 
with a pure Newtonian model on 5 Gyr revealed the opposite, and this system appeared far more unstable, with more than half of trajectories raising the  eccentricity of Mercury up to  0.9.

To confirm these results,  I have then proceeded to a direct integration, using a symplectic integrator (Laskar et al., 2004), of a pure  Newtonian planetary model, for 10  trajectories with close initial conditions. The result was consistent with the results of the secular system since 4 trajectories out of  10 led to eccentricity values for  Mercury larger than 0.9 (Laskar, 2008). This large excursion of the eccentricity of Mercury is explained by the presence of a resonance between the perihelion of Mercury and Jupiter, which is made easier in the absence of general relativity (GR). It is known that  GR increases the precession speed of the perihelion of mercury by $0.43''/yr$. This  moves it   from  $5.15''/yr$ to $5.58''/yr$, and thus send it further from  the value of the perihelion speed of Jupiter ($4.25''/yr$").
Independently, Batygin and Laughlin (2008) published similar results shortly after. The American team, which resumes the calculation of (Laskar, 1994) on a system of non-relativistic equations, also demonstrated the possibility of collisions between Mercury and Venus.
These results were still incomplete. Indeed, as the relativistic system is much more stable than the non-relativistic system, it is much more difficult to exhibit an orbit of collision between Mercury and Venus  in the realistic (relativistic) system  than in the non-relativistic system taken into consideration in these two previous studies. The real challenge was therefore in the estimation of the probability of collision of Mercury and Venus for a realistic, relativistic,  model. It is precisely this program that I had in mind since the writing of my  2008 paper, that allowed me  to estimate the probability of success of finding a collisional  orbit for  a realistic model of the Solar System.

\subsection{The search for Mercury-Venus collisional orbits}

With M. Gastineau, we then began a massive computation  of orbital solutions for the Solar System motion, under various aspects, the ambition being to confirm  and extend the results obtained 15 years before with the averaged equations. For this we used  a non-averaged model consistent with the short-term highly accurate INPOP planetary ephemeris that we had developed in the past years (Fienga et al., 2008).
Through the previous studies of  the secular system, I had estimated to  3 million hours the computing  time being necessary for such a study, but at the time,  no national computing center  was allowing even a 10 times smaller allocation of computing time.
We then used all the means which we could have access to: local cluster of workstations, computing center of  Paris Observatory, and a parallel machine that had just been installed  at  IPGP, Paris.  To search for additional CPU time, we also undertook the development of the first full-scale application  in  astronomy on the EGEE grid with 500 cores (Vuerli and al., 2009). These different runs, associated with multiple difficulties due to the variety of machines and operating systems, 
nevertheless  have allowed us to recover more than 2 million hours of CPU, but at the same time, a better estimate of the necessary computing time had  increased the required time to more than 5 million hours.
Quite fortunately, the availability of computing resources  has changed in France  in 2008, with the installation  of the JADE supercomputer at CINES, near Montpellier, with more than 12000  
cores\footnote{When this machine was installed, it was ranked 14 th worldwide among 
supercomputer centers.}.
As we could benefit of the experimental period on this machine, 
we started the computations as soon as the machine was switch on, 
in early August 2008, using 2501 cores, with one trajectory being computed on each core. 
We could then finalize our computations in about 6 months.

\subsection{Possibilité de collisions  entre Mercure, Mars, Vénus et la Terre}

\figi

With the JADE machine, we were able to simulate 2501 different solutions of the movement of the planets of the whole Solar System on 5 billion years, corresponding to the life expectancy of the system, before the Sun becomes a red giant.
The 2501 computed solutions are all compatible with our current knowledge of the Solar System. 
They should thus be considered as equiprobable outcomes of the futur of the Solar System.
In most of the solutions, the trajectories continue to evolve as in  the current few millions of years : the planetary orbits are deformed and precess under the influence of the mutual perturbations of the planets but without the possibility of collisions or ejections of planets outside the Solar System. Nevertheless, in 1\% 
of the cases, the eccentricity of Mercury increases considerably. 
In many cases, this deformation of the orbit of Mercury then leads to a collision with Venus, or
with  the Sun  in less than 5 Ga, while the orbit of the Earth remained little affected. 
However, for one of these orbits, the increase in the eccentricity of Mercury is followed by an increase in the eccentricity of Mars, and a complete internal destabilization of the inner Solar System (Mercury, Venus, Earth, Mars) in about 3.4 Gyr. 
Out of 201 additional cases studied in the vicinity of  this destabilization at about 3.4 Gyr, 5 ended by an ejection of Mars out of the Solar System. Others lead to collisions between the planets, or between a planet and the Sun in less than 100 million years.
One case resulted in a collision between Mercury and Earth, 29 cases in a collision between Mars and the Earth and 18 in a collision between Venus and the Earth (Laskar and Gastineau, 2009).
Beyond this spectacular aspect, these results  validate the methods of semi-analytical averaging developed for more than 20 years and which had allowed, 15 years ago, to show the possibility of collision between Mercury and Venus (Laskar, 1994). 

These results also answer to the question raised more than 300 years ago by Newton, by showing that collisions among planets or  ejections are actually possible within the life expectancy of the Sun, that is, in less than 5 Gyr. The main surprise that comes from the numerical simulations of the recent years is that the probability for this catastrophic events to occur is relatively high, of the order of 1\%, and thus not just a mathematical curiosity with extremely low  probability values. At the same time, 99\% of  the trajectories will behave in a similar way as in the recent past millions of years, 
which is coherent with our common understanding that the Solar System has not much evolved in the past 4 Gyr. What is more surprising is  that if we consider a pure Newtonian world,  the probability 
of  collisions within 5 Gyr grows to  60 \%, which can thus be considered as an additional 
indirect confirmation of general relativity.

\subsection*{Remerciements} 
This work has benefited from assistance from the Scientific Council of the Observatory of Paris, PNP-CNRS, and the computing center  of GENCI-CINES.

\def\bibl#1#2#3#4#5#6#7{{#2}:{\ #3}, {#4}, {\it #5}{\ \bf#6}{\
#7}}
\def\cel#1#2#3{{\it Celes. Mech. {\bf #1,}} #2-#3 }                


\begin{thebibliography}{aa}

\bibitem{} 
Applegate, J.H., Douglas, M.R., Gursel, Y., Sussman, G.J. and
Wisdom, J.: 1986, `The Solar System for 200 million years,' {\it
 Astron.\ J.} {\bf 92}, 176--194

\bibitem{}
\bibl{}{Arnold V.}{1963a}{Proof of Kolmogorov's theorem on the
preservation of quasi-periodic motions under small perturbations of the
hamiltonien}{Rus. Math. Surv.}{18,  N6} {9--36}

\bibitem{}
Arnold, V. I.: 1963b, `Small denominators and problems of stability of
motion in classical celestial mechanics,' {\it Russian  Math.
Surveys,} {\bf 18}, 6, 85--193 

\bibitem{BAT08}
Batygin, K., Laughlin, G. On the Dynamical Stability of the Solar System 
{\it ApJ}, {\bf 683}, 1207--1216 (2008)


\bibitem{} Brechenmacher, F., 2007, 
L'identité algébrique d'une pratique portée par la discussion sur l'équation 
à l'aide de laquelle on détermine les inégalités séculaires des planètes (1766-1874)
{\it Sciences et Techniques en Perspective}, IIe série, fasc. 1,  5--85

\bibitem{} 
\bibl{}{Bretagnon, P.}{1974}{Termes \`a longue
p\'eriodes dans le syst\`eme solaire} {Astron.
Astrophys}{30}{341--362}

\bibitem{} 
\bibl{}{Brumberg, V.A., Chapront, J.}{1973}{Construction of a general
planetary theory of the first order}{Cel. Mech.}{8}{335--355}


\bibitem{} 
\bibl{}{Carpino, M., Milani, A. and Nobili, A.M.}{1987}{Long-term numerical
integrations and synthetic theories for the motion of the outer
planets} {Astron.
Astrophys}{181}{182--194}


\bibitem{} \bibl{}{Cohen, C.J., Hubbard, E.C., Oesterwinter,
C.}{1973}{ } {Astron. Papers Am. Ephemeris}{XXII}{1}




\bibitem{} 
De la Lande, F. 1774, Abrégé d'Astronomie, première édition,
Paris

\bibitem{} 
De la Lande, F. 1795, Abrégé d'Astronomie, seconde édition, augmentée,
Paris

\bibitem{} 
Duriez, L.: 1979,  Approche d'une théorie générale planétaire en
variable elliptiques héliocentriques, {\it thèse} Lille



\bibitem{} Euler, L. 1752, Recherches sur les irrégularités du mouvement
de Jupiter et Saturne,


\bibitem{}
Féjoz, J., Herman, M.: 2004, 
Démonstration du  Théorème d'Arnold sur la stabilité du système planetaire (d'après Michael Herman), 
{\it Ergodic theory and Dynamical Systems}, {\bf 24}, 1--62

\bibitem{FIE08}
Fienga, A., Manche, H., Laskar, J., Gastineau, M., 2008,
INPOP06. A new numerical planetary ephemeris,
{\it A\&A}, {\bf 477}, {315--327}



\bibitem{} 
\bibl{}{ Giorgilli A.,  Delshams A.,  Fontich E.,  Galgani L.,  Simo
C.}{1989}{Effective stability for a Hamiltonian system near an elliptic equilibrium
point,  with an application to the restricted three body problem}{J.
Diff. Equa.}{77}{167--198}  


\bibitem{ITO02}
{Ito, T., Tanikawa, K.} {Long-term integrations and stability of planetary orbits in our Solar System},
{\it MNRAS}, {\bf 336}, {483--500} (2002)




\bibitem{} 
\bibl{}{Kinoshita, H., Nakai, H.}{1984}{Motions of the
perihelion of Neptune and Pluto}{Cel. Mech.}{34}{203}


\bibitem{} 
Kolmogorov, A.N.: 1954, On the conservation of conditionally periodic
motions under small perturbation of the Hamiltonian
{\it Dokl. Akad. Nauk. SSSR}, {\bf 98}, 469  


\bibitem{} Lagrange, 1766, Solution de différents problèmes de calcul
intégral, Miscellanea Taurinensia, t. III, 1762-1765,
{\it Oeuvres t. I, p. 471}




\bibitem{}
Lagrange, J. L.: 1776, Sur l'altération des moyens mouvements des planètes
{\it Mem. Acad.   Sci. Berlin}, 199  
{ Oeuvres complètes} {\bf VI} 255   Paris, Gauthier-Villars (1869)


\bibitem{} Lagrange, J.-L., 1778, Recherches sur les équations
séculaires des mouvements des n\oe uds et des inclinaisons des planètes,
Mémoires de l'Académie des Sciences de Paris, année 1774, publié en
1778.

\bibitem{} Lagrange, J.-L., 1781, Théorie des variations séculaires des
éléments  des planètes, Première partie Nouveaux Mémoires de l'Académie
des Sciences et Belles-Lettres de Berlin, année 1781. 
{\it \OE uvres, t. V, p. 125}


\bibitem{} Lagrange, J.-L., 1782, Théorie des variations séculaires des
éléments  des planètes, Seconde partie 
contenant la détermination de ces variations pour chacune des planètes principales, Nouveaux Mémoires de l'Académie
des Sciences et Belles-Lettres de Berlin, année 1782.
{\it \OE uvres, t. V, p. 211}

\bibitem{} Lagrange, J.-L., 1783a, Théorie des variations périodiques des 
mouvements des planètes, Première partie, Nouveaux Mémoires de l'Académie
des Sciences et Belles-Lettres de Berlin, année 1783. 
{\it \OE uvres, t. V, p. 347}

\bibitem{} Lagrange, J.-L., 1783b, Sur les variations séculaires des 
mouvements moyens des planètes,Nouveaux Mémoires de l'Académie
des Sciences et Belles-Lettres de Berlin, année 1783. 
{\it \OE uvres, t. V, p. 381}


\bibitem{} Lagrange, J.-L., 1784, Théorie des variations périodiques des 
mouvements des planètes, Seconde  partie, Nouveaux Mémoires de l'Académie
des Sciences et Belles-Lettres de Berlin, année 1784. 
{\it \OE uvres, t. V, p. 417}

\bibitem{}
 Lagrange, J.-L., 1808,
 Mémoire sur la théorie des variations des éléments des planètes 
 et en particulier des variations des grands axes de leurs orbites, 
Mémoires de la première classe de l'Institut de France, année 1808,
{\it\OE uvres, t. VI, p.713 }

\bibitem{}
 Lagrange, J.-L., 1808,
 Mémoire sur la théorie générale de la  variation des constantes arbitraires
dans tous les problèmes de la mécanique,  
 Mémoires de la première classe de l'Institut de France, année 1808,
{\it\OE uvres, t. VI, p.771}

\bibitem{}
 Lagrange, J.-L., 1809,
 Second Mémoire sur la théorie générale de la  variation des constantes arbitraires
dans les problèmes de mécanique,  
 Mémoires de la première classe de l'Institut de France, année 1809,
{\it\OE uvres, t. VI, p.809}


\bibitem{}
Laplace, P.S.: 1772, Mémoire sur les solutions particulières des équations
différentielles et sur les inégalités séculaires  des planètes
{Oeuvres complètes} {\bf 9} 325  Paris, Gauthier-Villars (1895)


\bibitem{} Laplace, P.-S., 1775, Mémoire sur les solutions particulières
des équations différentielles et sur les inégalités séculaires des
planètes. Mémoires de l'Académie des Sciences de Paris, année 1772,
publié en 1775, {\it\OE uvres, t. VIII, p. 325}



\bibitem{} Laplace, P.-S., 1776a, Sur le principe de la Gravitation
Universelle, et sur les inégalités séculaires des planètes qui en
dépendent, Mémoires de l'Académie des Sciences de Paris, , Savants étrangers, année 1773, t. VII, 
{\it\OE uvres, t. VIII, p. 201}

\bibitem{} Laplace, P.-S., 1776b, Mémoire sur l'Inclinaison moyenne des
orbites des comètes, sur la figure de la Terre et sur leur fonctions,
Mémoires de l'Académie des Sciences de Paris, Savants étrangers, année 1773, t. VII, 1776, 
{\it\OE uvres, t. VIII, p. 279}

\bibitem{}
Laplace, P.S.: 1784, Mémoire sur les inégalités séculaires des planètes et
des satellites
{\it Mem. Acad. royale des Sciences de Paris}, 
{Oeuvres complètes} {\bf XI} 49   Paris, Gauthier-Villars (1895)

\bibitem{} Laplace, P.S., 1785, Théorie de Jupiter et Saturne, Mémoires
de l'Académie Royale des Sciences de Paris, année 1785, 1788, 
{\it \OE uvres,  t. XI, p. 95}



\bibitem{laskar_theorie_1984}
Laskar, J. 1984: Théorie Générale Planétaire. Eléments orbitaux des planètes sur 1 million d'années, {\it Th{\`e}se}, Observatoire de Paris\\
\url{http://tel.archives-ouvertes.fr/tel-00702723}


\bibitem{}\bibl{}{Laskar, J.}{1985}{Accurate methods in general planetary
theory} {Astron. Astrophys.}{144}{133-146}


\bibitem{} 
\bibl{}{Laskar, J.}{1986}{Secular terms of classical planetary
theories using the results of general theory,}{Astron.
Astrophys.}{157}{59--70} 


\bibitem{} 
Laskar, J.:
1989, A numerical experiment on the chaotic behaviour of the Solar
System {\it Nature}, {\bf 338}, 237--238 
 
\bibitem{} 
Laskar, J.: 1990, The chaotic motion of the Solar System. A numerical
estimate of the size of the chaotic zones, {\it Icarus}, {\bf 88},
266--291


\bibitem{} 
Laskar, J.: 1992, La stabilit\'e du Syst\`eme Solaire, in {\it Chaos
et D\'eteminisme}, A. Dahan \etal, eds., Seuil, Paris 

\bibitem{} 
Laskar, J.: 2006, Lagrange et la stabilit\'e du Syst\`eme Solaire, in {\it Sfogliando 
La Mécanique Analitique}, Edizioni Universitarie di Lettere Economia Diritto, Milano,  pp. 157--174

\bibitem{LAS08}
Laskar, J.: 2008,
	Chaotic diffusion in the Solar System
	{\it Icarus}, {\bf 196}, 1--15

\bibitem{} 
Laskar, J., Quinn, T., Tremaine, S.: 1992a, Confirmation of Resonant
Structure in the Solar System, {\it Icarus}, {\it
95},{148--152}



\bibitem{}
\bibl{}{Laskar, J.}{1994}{Large scale chaos in the Solar System}{Astron.
Astrophys.}{287}{L9-L12}



 
 \bibitem{LAS04c} Laskar, J., Robutel, P., Joutel, F., Gastineau, M.,
Correia, A. C. M., Levrard, B.: 2004,
A long term numerical solution for
the insolation quantities of the Earth, 
{\it A\&A}, {\bf 428}, 261--285



\bibitem{M_LAS09c}
Laskar, J., Gastineau, M.: 2009,
Existence of colisional trajectories of Mercury, Mars and Venus 
with the Earth {\it Nature}, {\bf 459}, 817--819

\bibitem{Las11}
          Laskar, J., Fienga, A., Gastineau, M., Manche, H.: 2011,
       La2010: a new orbital solution for the long-term motion of the Earth, {\it A\&A}, {\bf 532}, A89



\bibitem{} Le Monnier, 1746a, Sur le Mouvement de Saturne et sur l'inégalité de ses 
révolutions périodiques, qui dépendent de ses diverses configurations à l'égard de Jupiter, 
première partie,   Mémoires de l'Académie Royale  des Sciences, Paris, 30 avril  1746, publié en 1751

\bibitem{} Le Monnier, 1746b, Sur le Mouvement de Saturne, Seconde Partie,   
Mémoires de l'Académie Royale  des Sciences, Paris, 7 mai 1746, publié en 1751

\bibitem{}
LeVerrier U.J.J.: 1840,
Mémoire sur les variations séculaires des éléments des orbites pour les sept planètes principales, Mercure, Vénus, la Terre, Mars, Jupiter, Saturne et Uranus, {\it Presented at the Academy of Sciences on September 16,  1839}, {Additions à la Connaissance des temps 
pour l'an 1843}  Paris, Bachelier,   pp 3--66.

\bibitem{}
LeVerrier U.J.J.: 1841,
Mémoire sur les inégalités  séculaires des   planètes, {\it Presented at the Academy of Sciences on December 14,  1840}, {Additions à la Connaissance des temps 
pour l'an 1844}  Paris, Bachelier,   pp 28--110.


\bibitem{} 
Lochak, P.: 1993,  Hamiltonian perturbation theory: periodic orbits, resonances
and intermittency, {\it Nonlinearity}, {\bf 6}, 885--904



\bibitem{} 
Morbidelli, A.,  Giorgilli, A.:  Superexponential stability of KAM tori, 
{\it J. Stat. Phys.} {\bf 78}, (1995) 1607--1617

\bibitem{} 
Moser, J.: 1962,
On invariant curves of area-preserving mappings of an annulus
{\it Nach. Akad. Wiss. Göttingen, Math. Phys. Kl.} II {\bf 1}, 1--20 


\bibitem{} 
Nekhoroshev, N.N.: 1977, An exponential estimates for the time of stability of
nearly integrable Hamiltonian systems, {\it Russian Math. Surveys}, {\bf 32},
1--65


\bibitem{} 
\bibl{}{Newhall, X. X., Standish, E. M., Williams, J. G.}{1983}{
DE102: a numerically integrated ephemeris of the Moon and planets spanning
forty-four centuries}
{Astron. Astrophys.}{125}{150--167}


\bibitem{} Newton, I., 1730, Opticks: Or, A Treatise of the Reflections,
Refractions, Inflexions and Colours of Light. The Fourth Edition,
corrected, 4th edition, London (seconde édition ``with Additions'' en
1717)


\bibitem{} Niederman, L., 1996, Stability over exponentially long times in the planetary problem,
{\it Nonlinearity} {\bf 9}, 1703--1751

\bibitem{} 
\bibl{}{Nobili, A.M., Milani, A. and Carpino, M.}{1989}{Fundamental
frequencies and small  divisors in the orbits of the outer planets}
{Astron. Astrophys.}{210}{313--336}

\bibitem{} 
 Poincar\'e, H.: 1892-1899, Les M\'ethodes Nouvelles de
la M\'ecanique C\'eleste, tomes I-III, {\it Gauthier Villard, Paris},
reprinted by Blanchard, 1987  

 
\bibitem{} 
 Poincar\'e, H.: 1897, Sur la stabilité du Système Solaire, {\it Annuaire du Bureau des Longitudes pour l'an 1898}, Paris, Gauthier-Villars,  B1-B16



\bibitem{}
Poisson, 1809,
 Mémoire sur les inégalités séculaires des moyens mouvments  des planètes,
lu à l'Académie le 20 juin 1808, Journal de l'Ecole Polytechnique,  Cahier XV, t. VIII, 1809, p.1--56



\bibitem{} 
Quinn, T.R., Tremaine, S., Duncan, M.: 1991, `A three million year
integration of the Earth's orbit,' {\it Astron.\ J.} {\bf 101},
2287--2305 

\bibitem{} 
Robutel, P.: 1995, Stability of the planetary three-body problem. II 
KAM theory and existence of quasiperiodic motions, \cel{62}{219}{261}


\bibitem{} 
Sussman, G.J., and Wisdom, J.: 1988, `Numerical evidence that the
motion of Pluto is chaotic.' {\it Science} {\bf 241},
433--437

\bibitem{} 
Sussman, G.J., and Wisdom, J.: 1992, 'Chaotic evolution of the solar
system', {\it Science} {\bf 257}, 56--62

\bibitem{} 
Wilson, C.: 1985, The great inequality of Jupiter and Saturn: from Kepler to Laplace, {\it Archive for History of Exact Sciences}, {\bf 33}, pp. 15–-290

\end{thebibliography}
\end{document}